# МАТЕМАТИЧНЕ МОДЕЛЮВАННЯ БІОХІМІЧНИХ ПРОЦЕСІВ

# SELF-OSCILLATORY DYNAMICS OF THE METABOLIC PROCESS IN A CELL

*V. I. GRYTSAY[1], I. V. MUSATENKO[2]*

[1]*Bogolyubov Institute for Theoretical Physics,*
*National Academy of Sciences of Ukraine, Kyiv;*
*e-mail: vgrytsay@bitp.kiev.ua;*
[2]*Taras Shevchenko National University of Kyiv, Ukraine;*
*e-mail: ivmusatenko@gmail.com*

*In this work, a mathematical model of self-oscillatory dynamics of the metabolism in a cell is studied. The full phase-parametric characteristics of variations of the form of attractors depending on the dissipation of a kinetic membrane potential are calculated. The bifurcations and the scenarios of the transitions «order-chaos», «chaos-order» and «order-order» are found. We constructed the projections of the multidimensional phase portraits of attractors, Poincaré sections, and Poincaré maps. The process of self-organization of regular attractors through the formation torus was investigated. The total spectra of Lyapunov exponents and the divergences characterizing a structural stability of the determined attractors are calculated. The results obtained demonstrate the possibility of the application of classical tools of nonlinear dynamics to the study of the self-organization and the appearance of a chaos in the metabolic process in a cells.*

K e y  w o r d s: *mathematical model, metabolic process, self-organization, strange attractor, Poincaré section, Lyapunov exponents.*

The development of a united theory of self-organization of the matter is one of the most important problems of natural sciences. Its solution is possible only at the comprehension of the unity of the physical laws for alive and inorganic matters. It is worth noting the significant contribution of synergetics to this field of knowledge. The idea of the appearance of dissipative structures in any open nonlinear systems is a corner stone in the foundation of such a theory. Synergetics explains the formation of an order from a chaos and, by this, outlines a way to the description of the biochemical evolution of the inorganic matter to the alive one [1–4].

The development of a general mathematical model of cell is a very complicated task. To solve it, it is necessary to study the biochemical processes and the types of nonlinearities characteristic of enzyme-substrate interactions causing the appearance of the self-organization in a system. Namely this problem belongs to the field of interests of the author.

Here, we deal with a mathematical model of the metabolism running in a cell *Arthrobacter globiformis*. This model of enzyme kinetics is constructed on the basis of the original method developed by V. P. Gachok [5, 6] who proposed to apply a function of the general form $V(X) = aX/(1 + bX)$ describing the adsorption of an enzyme in the domain of local coupling with a substrate to the modeling of biochemical processes.

As was noticed by a number of researchers [7, 8], Prof. V. P. Gachok unified the construction of models for a wide class of biochemical systems described with multidimensional nonlinear differential equations.

In the present work, we will consider the metabolic process in the *Arthrobacter globiformis* cells quite well studied in experiments [9]. Bacteria of the given species are widely spread in the Nature and intensively used as producers of various metabolites. These bacteria are applied to the purification of wastes of the petrochemical, cokechemical, tannic, etc. productions, because they are able to decompose practically all hydrocarbons contained in oil. They are used to eliminate the toxic action of herbicides on plants and in other fields of ecology. In addition, these bacteria are involved in medicine in the diagnostics of the galactose exchange and in the transformation of steroids. Thus, the significance of *A. globiformis* culture for the industry is obvious. It is also worth noting that its cultivation in the continuous mode is characterized by the appearance of oscillatory phenomena in the metabolism of the given cells [10, 11], which should be taken into account in their usage.

Such oscillatory mode was theoretically forecasted in work [12] prior to the experiment. Analogous modes of metabolic self-oscillations were registered in the processes of photosynthesis, glycolysis, variation of calcium in cells, oscillations in heart muscle, etc. [13–16]. Namely the oscillatory modes allow one to study various structural-functional couplings, which lead to the self-organization of metabolic processes in cells and a whole biosystem.

Earlier, a mathematical model of the process of transformation of steroids in a flow bioreactor was developed jointly with experimenters [17, 18]. In the biochemical processes, *Arthrobacter globiformis* cells immobilized in granules were used. The executed numerical experiment revealed the oscillatory modes [19–30].

As distinct from the earlier results obtained within the model under consideration, we will calculate phase-parametric characteristics of variations of the form of attractors depending on the dissipation of a kinetic membrane potential and will show the appearance of the mutual transitions between stable and chaotic oscillations. With the help of a mathematical apparatus of the nonlinear dynamics, we will perform the investigation of these transitions in the mentioned modes. The calculated scenarios such as «order-chaos», «chaos-order» and «order-order» and their mathematical analysis give a certain contribution to the understanding of the self-organization of biochemical systems.

### A mathematical model and methods of its study

The general scheme of metabolic process running in *Arthrobacter globiformis* cells under the transformation of steroids is presented in Fig. 1. The input substances for a bioreactor are hydrocortisone and oxygen. The output products are prednisolone and its hydroxy derivative. The process occurs in the following way.

Hydrocortisone ($G$) present in the flowing solution induces the biosynthesis of 3-ketosteroid-$\Delta'$-dehydrogenase enzyme ($E_1$) on the external side of a cytoplasmatic membrane of cells. Under the action of this enzyme, hydrocortisone is transformed into prednisolone ($P$). The 3-ketosteroid-$\Delta'$-dehydrogenase enzyme acquires the reducing form ($e_1$). By reducing the respiratory chain ($q$), this en-zyme is oxidized into ($E_1$). Oxygen ($O_2$) from the flowing solution oxidizes the respiratory chain into the form ($Q$). Produced prednisolone induces the biosynthesis of 20β-oxysteroid-dehydrogenase enzyme ($E_2$), which together with $NAD·H(N)$ coenzyme produces 20β-oxyderivative of prednisolone ($B$). Then a part of produced 20β-oxyderivative of prednisolone is consumed by cells in the Krebs cycle, by increasing the level of $NAD·H$. In the enhanced concentration, $NAD·H$ blocks the respiratory chain. In this case, the process of transformation of hydrocortisone becomes slow.

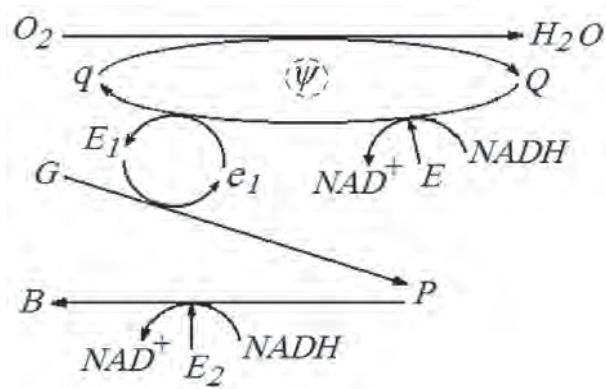

*Fig. 1. General scheme of the metabolic process of Arthrobacter globiformis cell*

The transformation of steroids is accompanied by the formation of a kinetic membrane potential ($\psi$). Its increased level holds the respiratory chain in the reduced state ($q$). The membrane potential is also used by cells in the Krebs cycle.

The mathematical model of the process under flow conditions of a bioreactor is constructed in accordance with the general scheme (Fig. 1) of metabolic processes running in *Arthrobacter globiformis* cells under the transformation of steroids [19–30]:

$$\frac{dG}{dt} = \frac{G_0}{N_3 + G + \gamma_2\psi} - l_1 V(E_1)V(G) - \alpha_3 G, \quad (1)$$

$$\frac{dP}{dt} = l_1 V(E_1)V(G) - l_2 V(E_2)V(N)V(P) - \alpha_4 P, \quad (2)$$

$$\frac{dB}{dt} = l_2 V(E_2)V(N)V(P) - k_1 V(\psi)V(B) - \alpha_5 B, \quad (3)$$

$$\frac{dE_1}{dt} = E_{10} \frac{G^2}{\beta_1 + G^2}(1 - \frac{P + mN}{N_1 + P + mN}) -$$
$$- l_1 V(E_1)V(G) + l_4 V(e_1)V(Q) - \alpha_1 E_1, \quad (4)$$

$$\frac{de_1}{dt} = -l_4 V(e_1)V(Q) + l_1 V(E_1)V(G) - \alpha_1 e_1, \quad (5)$$

$$\frac{dQ}{dt} = 6lV(2-Q)V(O_2)V^{(1)}(\psi) - l_6 V(e_1)V(Q) -$$
$$- l_7 V(Q)V(N), \qquad (6)$$

$$\frac{dO_2}{dt} = \frac{O_{20}}{N_5 + O_2} - lV(2-Q)V(O_2)V^{(1)}(\psi) - \alpha_7 O_2, \quad (7)$$

$$\frac{dE_2}{dt} = E_{20}\frac{P^2}{\beta_2 + P^2}\frac{N}{\beta + N}(1 - \frac{B}{N_2 + B}) -$$
$$- l_{10} V(E_2)V(N)V(P) - \alpha_2 E_2, \qquad (8)$$

$$\frac{dN}{dt} = -l_2 V(E_2)V(N)V(P) - l_7 V(Q)V(N) +$$
$$+ k_2 V(B)\frac{\psi}{K_{10} + \psi} + \frac{N_0}{N_4 + N} - \alpha_6 N, \qquad (9)$$

$$\frac{d\psi}{dt} = l_5 V(E_1)V(G) + l_8 V(N)V(Q) - \alpha\psi. \qquad (10)$$

Here, $V(X) = X/(1 + X)$; $V^{(1)}(\psi) = 1/(1+\psi^2)$; $V(X)$ is a function related to the adsorption of the enzyme in the domain of local coupling (it was proposed by V. P. Gachok [5, 6]); and $V^{(1)}(\psi)$ is a function characterizing the influence of the kinetic membrane potential on the respiratory chain.

The reduction of the variables of the system to the dimensionless form is given in works [17, 18]:

$$G = \frac{[G]}{K_G}; \quad P = \frac{[P]}{K_P}; \quad B = \frac{[B]}{K_B}; \quad E_1 = \frac{[E_1]}{K_{E_1}};$$

$$e_1 = \frac{[e_1]}{K_{e_1}}; \quad Q = \frac{[Q]}{K_Q}; \quad O_2 = \frac{[O_2]}{K_{O_2}}; \quad E_2 = \frac{[E_2]}{K_{E_2}};$$

$$N = \frac{[N]}{K_N}; \quad \psi = \frac{[\psi]}{K_\psi}.$$

The parameters of the system are as follows: $l = l_1 = k_1 = 0.2$; $l_2 = l_{10} = 0.27$; $l_5 = 0.6$; $l_4 = l_6 = 0.5$; $l_7 = 1.2$; $l_8 = 2.4$; $k_2 = 1.5$; $E_{10} = 3$; $\beta_1 = 2$; $N_1 = 0.03$; $m = 2.5$; $\alpha = 0.033$; $a_1 = 0.007$; $\alpha_1 = 0.0068$; $E_{20} = 1.2$; $\beta = 0.01$; $\beta_2 = 1$; $N_2 = 0.03$; $\alpha_2 = 0.02$; $G_0 = 0.019$; $N_3 = 2$; $\gamma_2 = 0.2$; $\alpha_5 = 0.014$; $\alpha_3 = \alpha_4 = \alpha_6 = \alpha_7 = 0.001$; $O_{20} = 0.015$; $N_5 = 0.1$; $N_0 = 0.003$; $N_4 = 1$; $K_{10} = 0.7$.

The parameters were obtained using comparison of experimental and computed characteristics of a stationary regime in the flow bioreactor. [17, 18].

Equations (1)−(9) describe variations of the concentrations: (1) − hydrocortisone ($G$); (2) − prednisolone ($P$); (3) − 20β-oxyderivative of prednisolone ($B$); (4) − oxidized form of 3-ketosteroid-Δ'-dehydrogenase ($E_1$); (5) − reduced form of 3-ketosteroid-Δ'-dehydrogenase ($e_1$); (6) − oxidized form of the respiratory chain ($Q$); (7) − oxygen ($O_2$); (8) − 20β-oxysteroid-dehydrogenase ($E_2$); (9) − $NAD·H$ (reduced form of nicotinamideadeninedinucleotide) ($N$). Equation (10) describes the variation of the kinetic membrane potential ($\psi$).

This autonomous system of nonlinear differential equations was numerically solved by the Runge-Kutta-Merson method. The accuracy of calculations was set to be $10^{-8}$. To ensure the reliability of the study, namely the transition of the system from the initial and transient phases to the asymptotic state with an attractor, the computation time was taken to be 1 000 000. For this time, the trajectory was «stuck» onto the corresponding attractor.

The given system is dissipative. It is characterized by the dissipation of the concentrations of its reagents (except for the respiratory chain). Therefore, the elements of its phase space shrink during the evolution. Let us denote the set of all initial points of the system by $V$: $\vec{x}_0 \in V$, let $\vec{x}(t) \in L$ as $t \to +\infty$ and $L \in V$. The limiting set $L$ is an attracting one, and the system approaches an attractor, so that $V$ is called the basin of a pulling attractor. We have obtained the following attractors: stationary modes (stable focus and stable node), self-oscillatory modes with various multiplicities and strange attractors.

We analyzed the stability of modes of the metabolic process with the use of the theory of stability of solutions of differential equations [31, 32], which was already applied to the study of many-stage enzymic reactions [33, 34].

For the unambiguous identification of the type of an attractor, we calculated the total spectrum of Lyapunov exponents, for chosen points, $\lambda_1, \lambda_2,...,\lambda_{10}$ and determined their sum: $\Lambda = \sum_{j=1}^{10} \lambda_j$ (see Table). The calculation was carried out by Benettin's algorithm. The orthogonalization of the perturbation vectors was performed within the Gram−Schmidt method [31, 35, 36].

The studies within the model were carried out by the construction of phase-parametric characteristics, by changing one of the parameters (see below).

In order to identify the structure of attractors at the chosen points, we used the method of Poincaré sections. We note that the Poincaré section possesses properties of a flow and can be studied easier than the flow itself. In the phase space, we drew the cutting plane $P = 0.2$ and determined the coordinates of intersections for the corresponding variables on the two-dimensional projections of the constructed section. By the positions of the intersection points in this plane, we can judge about the structure of an attractor and the topological configuration of its phase trajectory.

*Total spectra of Lyapunov exponents for attractors of the system under study ($\lambda_4$-$\lambda_9$ are not important for our investigation)*

| $\alpha$ | Attractor | $\lambda_1$ | $\lambda_2$ | $\lambda_3$ | $\lambda_4$-$\lambda_9$ | $\lambda_{10}$ | $\Lambda$ |
|---|---|---|---|---|---|---|---|
| 0.032 | $1 \cdot 2^0$ | .000048 | -.000455 | -.000329 | --- | -.506441 | -.891283 |
| 0.03211 | $1 \cdot 2^0$ (t) | .000049 | -.000026 | .000090 | --- | -.519101 | -.902325 |
| 0.0321107 | $8 \cdot 2^0$ | .000073 | -.000807 | -.000797 | --- | -.508550 | -.898564 |
| 0.0321125 | $\approx 1 \cdot 2^0$ (t) | .000045 | -.000011 | .000100 | --- | -.519370 | -.902561 |
| 0.03211295 | $8 \cdot 2^0$ | .000278 | .000036 | -.002045 | --- | -.507725 | -.898162 |
| 0.03212 | $14 \cdot 2^0$ | .000016 | -.000203 | -.002977 | --- | -.496171 | -.896993 |
| 0.0321286 | $\approx n \cdot 2^0$ (t) | .000055 | .000018 | .000065 | --- | -.520426 | -.904159 |
| 0.032135 | $47 \cdot 2^0$ (t) | .000056 | .000035 | -.000007 | --- | -.520343 | -.904551 |
| 0.032136 | $\approx 39 \cdot 2^0$ (t) | .000059 | .000026 | -.000013 | --- | -.520382 | -.904657 |
| 0.032137 | $39 \cdot 2^0$ (t) | .000052 | .000048 | -.000028 | --- | -.520452 | -.904776 |
| 0.032138 | $70 \cdot 2^0$ (t) | .000057 | .000041 | -.000025 | --- | -.520497 | -.904883 |
| 0.032139 | $14 \cdot 2^0$ | .000042 | -.000212 | -.003185 | --- | -.497477 | -.898991 |
| 0.0321395 | $14 \cdot 2^0$ | .000039 | -.000191 | -.003196 | --- | -.497398 | -.898906 |
| 0.0321496 | $14 \cdot 2^0$ | .000043 | -.000539 | -.002878 | --- | -.496308 | -.898024 |
| 0.0321596 | $14 \cdot 2^0$ | .000040 | -.000142 | -.003306 | --- | -.496728 | -.898550 |

Using the obtained Poincaré sections, we constructed the Poincaré maps, which allow one to determine the stationary mode of oscillations in the system.

### Results of studies

System (1) − (10) is a completely determined nonlinear system of differential equations in the ten-dimensional phase space. It describes the biochemical system with positive feedback, which generates the autocatalysis of *NAD·H* (see Eqs. (3) and (9)). Therefore, the self-oscillatory modes can appear in it. With the help of numerical computations, we have studied all possible modes arising in the given system at a variation of the dissipation of the kinetic membrane potential $\alpha$ (10). This potential is the most general parameter determining the stability of the metabolic process in a cell.

In Fig. 2,*a*, we present a phase-parametric characteristic of the system $G(\alpha)$ for the interval, where $\alpha \in (0.032, 0.033)$. To construct the phase-parametric characteristic, we used the method of sections. In the phase space with a trajectory of the system, we drew the cutting plane $P = 0.2$. Such a choice is supported by the symmetry of oscillations relative to this point in multiple modes. As an example, Fig. 2,*b,c* shows the kinetics of $G(t)$ and $P(t)$ and the projections of phase portraits, which are used in the construction of the phase-parametric characteristic. When the trajectory $P(t)$ intersects this plane in some single direction, we mark the value of chosen variable (e.g., $G$) on the phase-parametric characteristic. In such a way, we determine the point of the intersection of the trajectory with the two-dimensional plane. If a multiple periodic limiting cycle arises, we obtain a number of points on the plane, which are repeated in the period. In the case where a deterministic chaos appears, the points of intersection of the trajectory with the plane are located chaotically.

In Fig. 2,*a*, we present the change of the multiplicity of oscillations in various modes for the whole interval of variation of $\alpha$. There, the regions of the formation of regular and chaotic attractors are clearly seen. By examining the figure from right to left, i.e., by decreasing $\alpha$ from 0.033 down to 0.032, we see the appearance of the sequence of regular $n \cdot 2^0$ and strange $n \cdot 2^x$ attractors, whose cycle period corresponds to the multiplicity $n$. If the given attractor is formed on a toroidal surface, it is denoted by the mark (*t*) (Table).

The less the coefficient of dissipation of the kinetic membrane potential of a cell, the more the energy is conserved in it, which complicates its internal dynamics. The 8-fold periodic cycle of a regular attractor $8 \cdot 2^0$ seen at $\alpha = 0.033$ is broken with the formation of a strange attractor $8 \cdot 2^x$. In other words, the transition «order-chaos» occurs. The further decrease in $\alpha$ causes the self-organization of the system and the formation of a 9-fold periodic cycle, which corresponds to the transition «chaos-order». The given regular attractor $9 \cdot 2^0$ is replaced by a strange attractor $9 \cdot 2^x$, then by $10 \cdot 2^0$, $10 \cdot 2^x$, and so on up to the regular attractor of a quasiperiodic cycle $\approx n \cdot 2^0$. On this section of the phase-parametric characteristic, the decrease in the coefficient of dissipation of the kinetic membrane potential is periodically accompanied by the regular self-organization of the system. There occur the destruction of simple structures and the appearance of more complex ones. But, after the attractor $\approx n \cdot 2^0$, a simple regular attractor $1 \cdot 2^0$ appears unexpectedly. Let us consider the specific features of the self-organization of the system on separate intervals of the phase-parametric trajectory in more details.

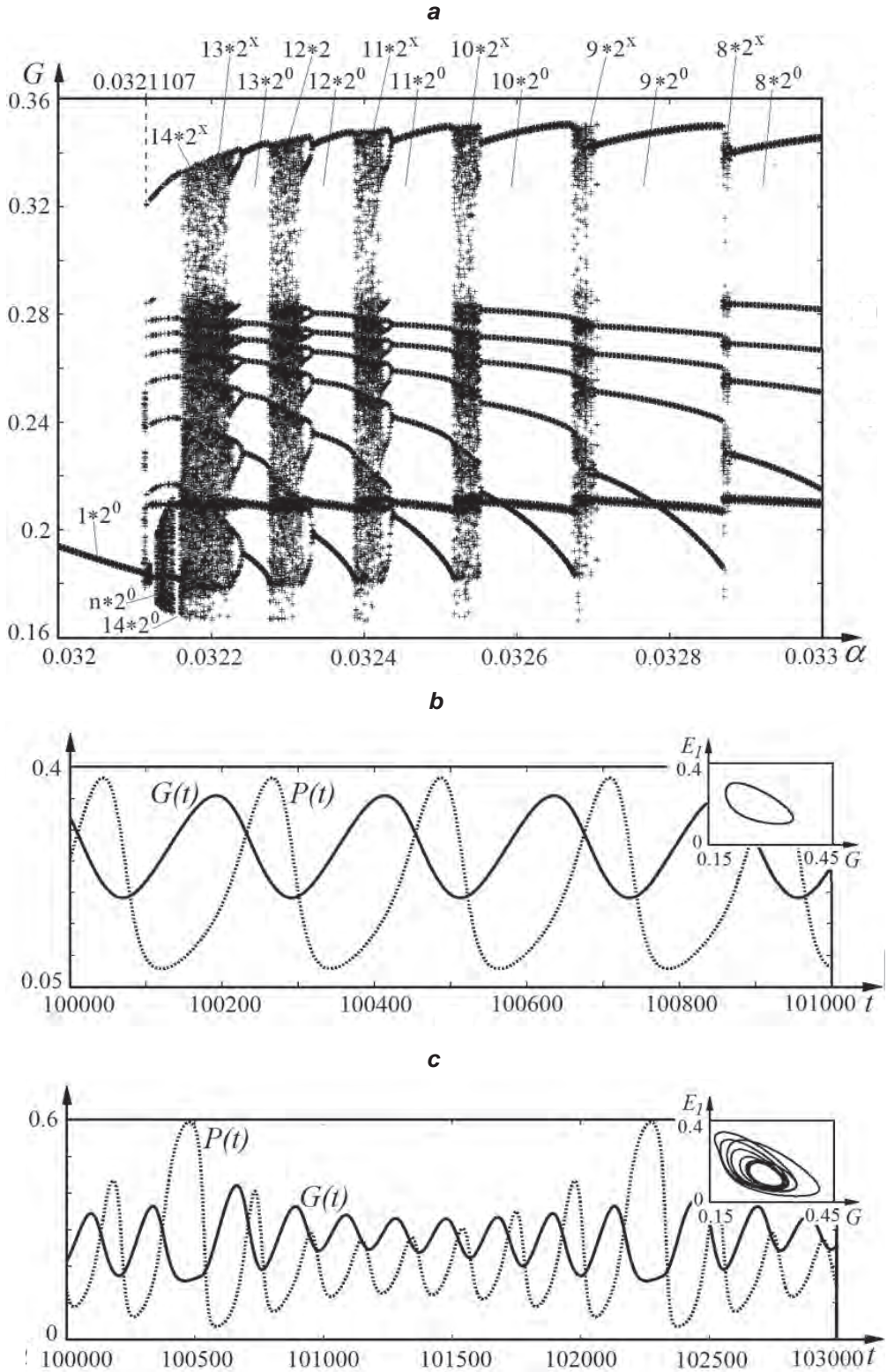

*Fig. 2. (a) Phase-parametric characteristic of the system for $\alpha \in (0.032, 0.0321107)$; (b) The kinetic curves $G(t)$ and $P(t)$ with projections of the phase portrait for a 1-fold periodic mode for $\alpha = 0.032$; (c) 8-fold periodic mode for $\alpha = 0.0321107$*

In Fig. 2,b,c, we present the kinetic behavior of the variables $G(t)$ and $P(t)$ and the projections of phase portraits for $\alpha = 0.032$ and $\alpha = 0.0321107$, respectively. In the interval $\alpha \in (0.032, 0.03211)$, the system transits from a simple 1-fold periodic cycle to a 1-fold periodic cycle on a torus (Table). In this case, the amplitude increases.

At $\alpha = 0.0321107$, a simple 8-fold periodic cycle appears instantly from the 1-fold cycle without intermediate doubling. Its kinetics and phase portrait are shown in Fig. 2,c. In other words, we observe the instantaneous transition «order-order». The found cycles are stable by Lyapunov, because their first Lyapunov exponents are equal to zero (Table).

By analyzing this transition, we note that the divergence of the system $\Lambda$ is negative at the intermediate point $\alpha = 0.03211$ and is larger in modulus than that at $\alpha = 0.032$ and $0.0321107$. For this mode, $\lambda_1$, $\lambda_2$, and $\lambda_3$ are zero. But, for the modes with $\alpha = 0.032$ and $0.0321107$, $\lambda_2$ and $\lambda_3$ are negative. This means that the element of the phase volume of the system at these points shrinks, as a whole, more strongly along the trajectory, i.e., the trajectory moves away from the toroidal surface ($\alpha = 0.03211$) onto simple regular attractors.

The increase of $|\Lambda|$ at $\alpha = 0.03211$ is regular, because the dissipation ($\alpha$) increases. As $\alpha$ increases up to $0.0321107$, $|\Lambda|$ decreases. This means that, after the appearance of a bifurcation and the transition from $1 \cdot 2^0$ to $8 \cdot 2^0$, the metabolism in a cell organizes itself so that the total dissipation of the cell energy $|\Lambda|$ becomes less at a higher dissipation of the kinetic membrane potential $\alpha$. We note that the energy dissipation under oscillations realized on a toroidal surface in the multidimensional space is higher than that under simple periodic oscillations.

As $\alpha$ increases, the formed cycles are slowly shifted, on the whole, in the phase space in the single direction. It is seen in Fig. 2,a from inclined displacements of the traces of intersection points.

Let us now consider each part of the phase-parametric characteristic in more details.

In Fig. 3,a, we show a part of the phase-parametric characteristic for $\alpha \in (0.03211, 0.03215)$. As $\alpha$ increases to $0.03212$, we observe the alternation of attractors. For example, a one-fold quasiperiodic cycle $\approx 1 \cdot 2^0$ arises again on a toroidal surface at $\alpha = 0.0321125$. Then this cycle is broken at $\alpha = 0.03211295$, and a 8-fold strange attractor $8 \cdot 2^x$ arises. For this attractor, a part of the projection of its phase portrait is shown in Fig. 3,b. At $\alpha = 0.03212$, the oscillations become simple, and an ordinary 14-fold periodic cycle is established (Fig. 3,c). As $\alpha$ increases to $0.0321286$, the oscillations transit onto a toroidal surface. Quasiperiodic oscillations $\approx n \cdot 2^0$ with irrational ratios of their frequencies are observed. The initial stage of the filling of the phase space by the trajectory is demonstrated in Fig. 3,d by a part of the projection of the phase portrait.

As $\alpha$ increases further to $0.032135$ (Fig. 3,a), we see the formation of lacunes on the phase-parametric characteristic, which means that the given multiple quasiperiodic attractors $\approx n \cdot 2^0$ alternate with simple periodic 14-fold attractors. At $\alpha = 0.032135$ as a result of the superposition of two periodic multiple oscillations with rationally commensurable frequencies, the strictly periodic mode $47 \cdot 2^0$ of oscillations on a torus is established. Its part is shown in Fig. 3,e.

In Fig. 4,a, we present the kinetic curve of this mode for the variable $e_1(t)$ (most clearly pronounced). The plot shows the periodic repetition of stable 47-fold oscillations.

The further increase in the parameter $\alpha$ leads to a change in the periodicity of oscillations on the toroidal surface. The quasiperiodic oscillations $\approx 39 \cdot 2^0$ are established at $\alpha = 0.032136$, strictly periodic oscillations $39 \cdot 2^0$ at $\alpha = 0.032137$, and then periodic 70-fold oscillations at $\alpha = 0.032138$. At $\alpha = 0.032139$, the oscillations become simple 14-fold. Their kinetics for the variable $e_1(t)$ ($\alpha = 0.0321395$) is shown in Fig. 4,b, and the projection of the phase portrait of the regular attractor $14 \cdot 2^0$ is analogous to that in Fig. 3,c.

On the considered interval of variation of $\alpha$, we observe the alternation of attractors with different complexities and the transitions: «order-order», «order-chaos», and «chaos-order». At $\alpha = 0.0321496$, the alternation stops (Fig. 3,a). Up to $\alpha = 0.0321596$, we see the modes of simple periodic oscillations $14 \cdot 2^0$.

In Fig. 5, we constructed the projections of a Poincaré map by the plane $P = 0.2$ for, respectively, the regular attractor of a simple periodic cycle $8 \cdot 2^0$ ($\alpha = 0.0321107$) (a); the regular attractor of a quasiperiodic cycle $\approx 1 \cdot 2^0$ on a toroidal surface ($\alpha = 0.0321125$) (b); the regular attractor

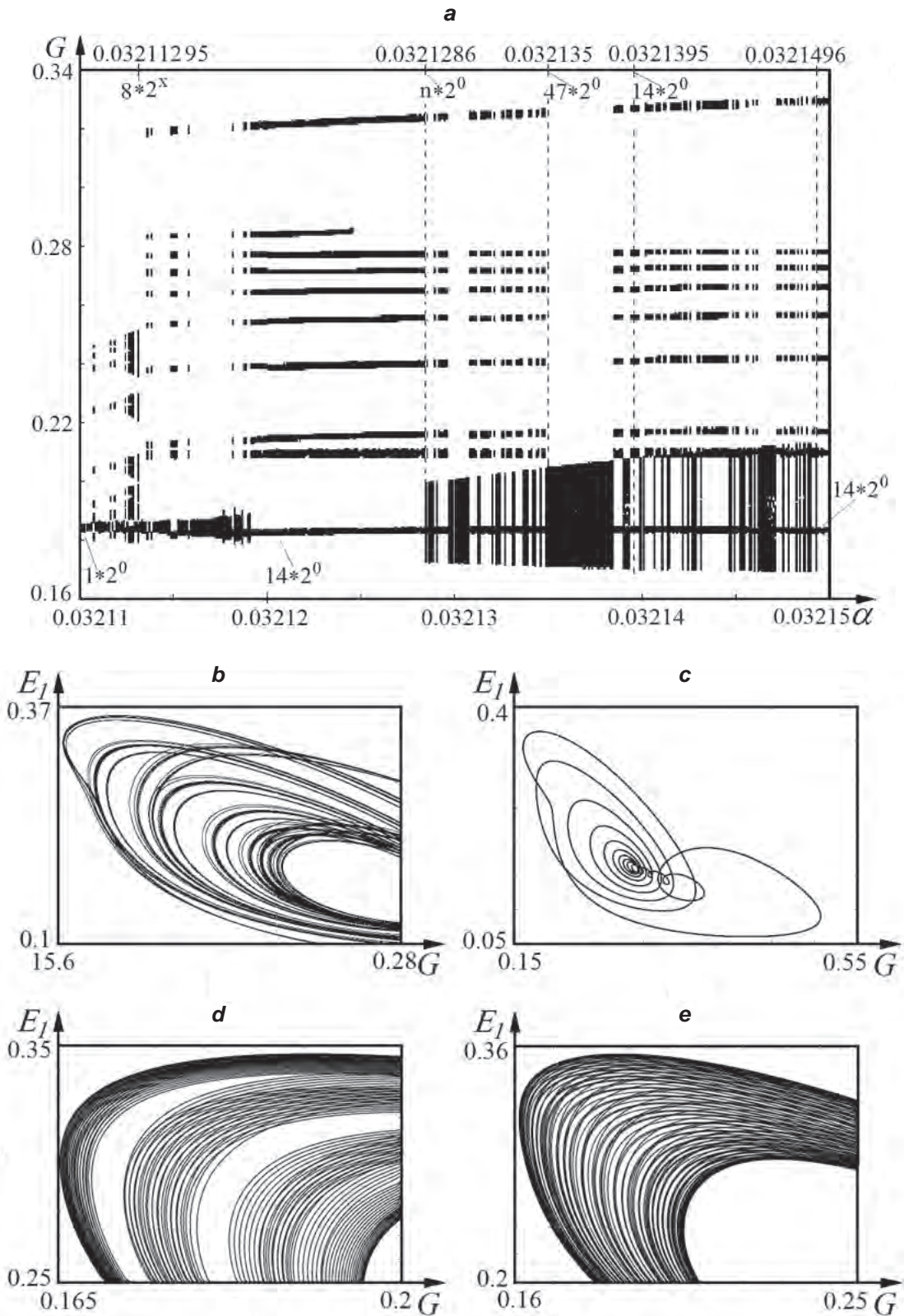

*Fig. 3. (a) Phase-parametric characteristic of the system at α∈(0.03211, 0.03215); (b) parts of the projections of phase portraits for α = 0.03211295; (c) parts of the projections of phase portraits for α = 0.03212; (d) parts of the projections of phase portraits for α = 0.0321286; (e) parts of the projections of phase portraits for α = 0.032135*

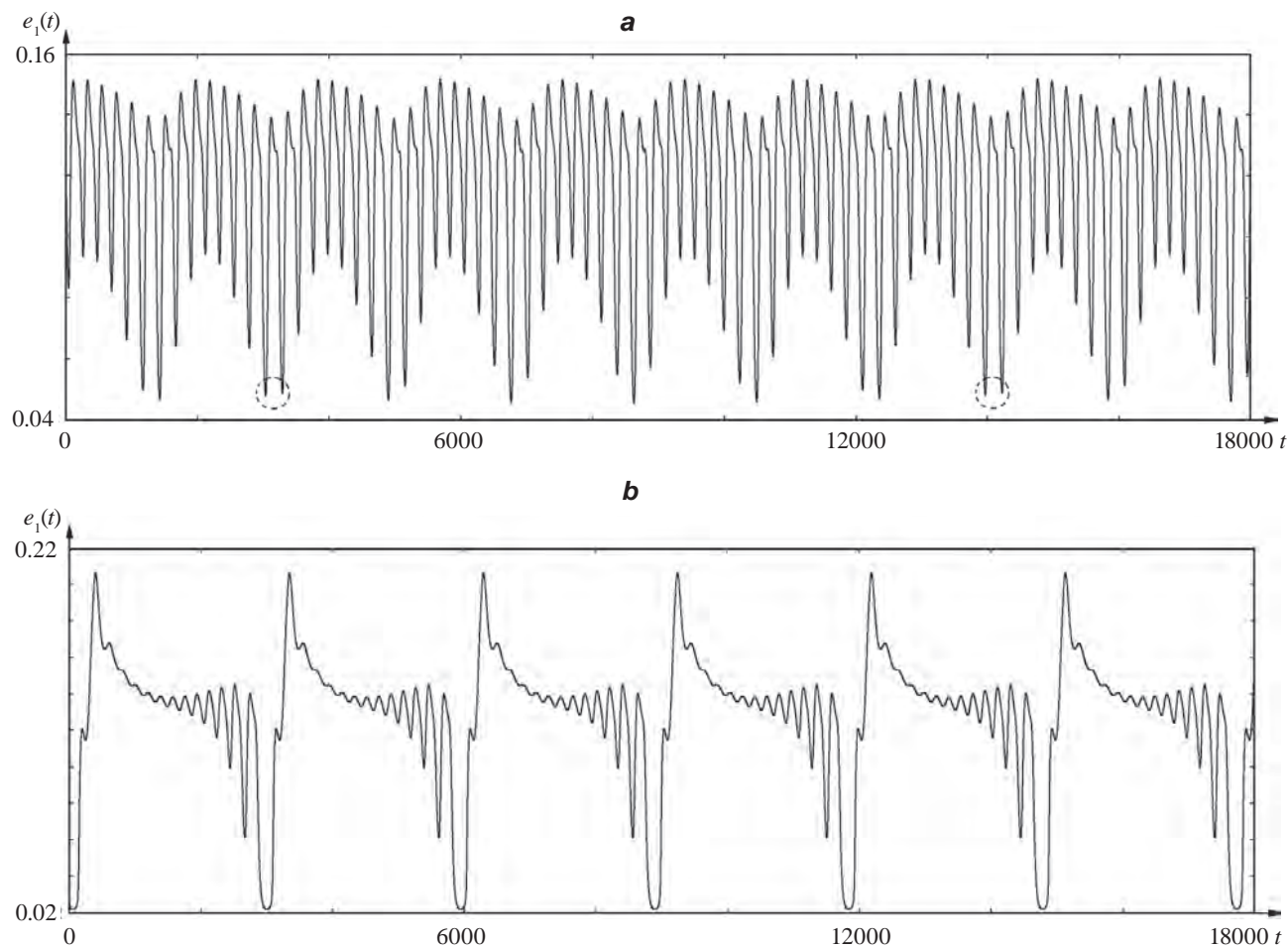

*Fig. 4. (a) Kinetic curves of the variable $e_1(t)$ of regular attractors $47 \cdot 2^0$ at $\alpha = 0.032135$; (b) regular attractors $14 \cdot 2^0$ at $\alpha = 0.0321395$*

of a quasiperiodic cycle $\approx n \cdot 2^0$ on a toroidal surface ($\alpha = 0.0321286$) (c), and the regular attractor of a periodic cycle $47 \cdot 2^0$ on a toroidal surface ($\alpha = 0.032135$) − (d).

The projections correspond to modes, whose type is determined by the calculated Lyapunov exponents. The location of the points of intersection of the plane by a trajectory corresponds to the topological configuration of the trajectory.

In Fig. 5,a, we show the intersection points for a 8-fold periodic cycle. These points are strongly repeated in the period of the given cycle. Figure 5,b presents the set of the points of intersection of the plane by a quasiperiodic cycle $\approx 1 \cdot 2^0$. Their maximum density is observed in the central part, which is a region of attraction of a onefold cycle. In Fig. 5,c,d, we give, respectively, the Poincaré sections for modes in the case where the phase trajectory moves along the torus surface. In this case, the phase trajectory on the torus surface can be considered as a superposition of two motions: a rotation around «parallels» of the torus with frequency $f_1$ and a rotation around the «cylinder» forming the torus with frequency $f_2$. The points of intersection of the plane $P = 0.2$ by the phase trajectory arise in regular time intervals equal to the period $T = 1/f_1$. The shape of the Poincaré section depends on the ratio $f_1/f_2$. At $\alpha = 0.0321286$, this ratio is irrational. Therefore, the phase trajectory is closed and covers the torus surface everywhere densely. The regular attractor of a quasiperiodic cycle $\approx n \cdot 2^0$ is formed. The Poincaré section is a closed curve in the form of an ellipse (Fig. 5,c). At $\alpha = 0.032135$, the ratio $f_1/f_2 = n_1/n_2$ is rational, and the Poincaré section consists of a finite number of points, which are located on a closed curve. In this case, the periodic motion with the period $T = n_1/f_1 = n_2/f_2$ occurs.

In other words, the phase trajectory is closed after $n_1$ rotations around «parallels» and $n_2$ rota-tions around «meridians» with the formation of a regular attractor 47·2⁰. The Poincaré section contains 47 points (Fig. 5,*d*).

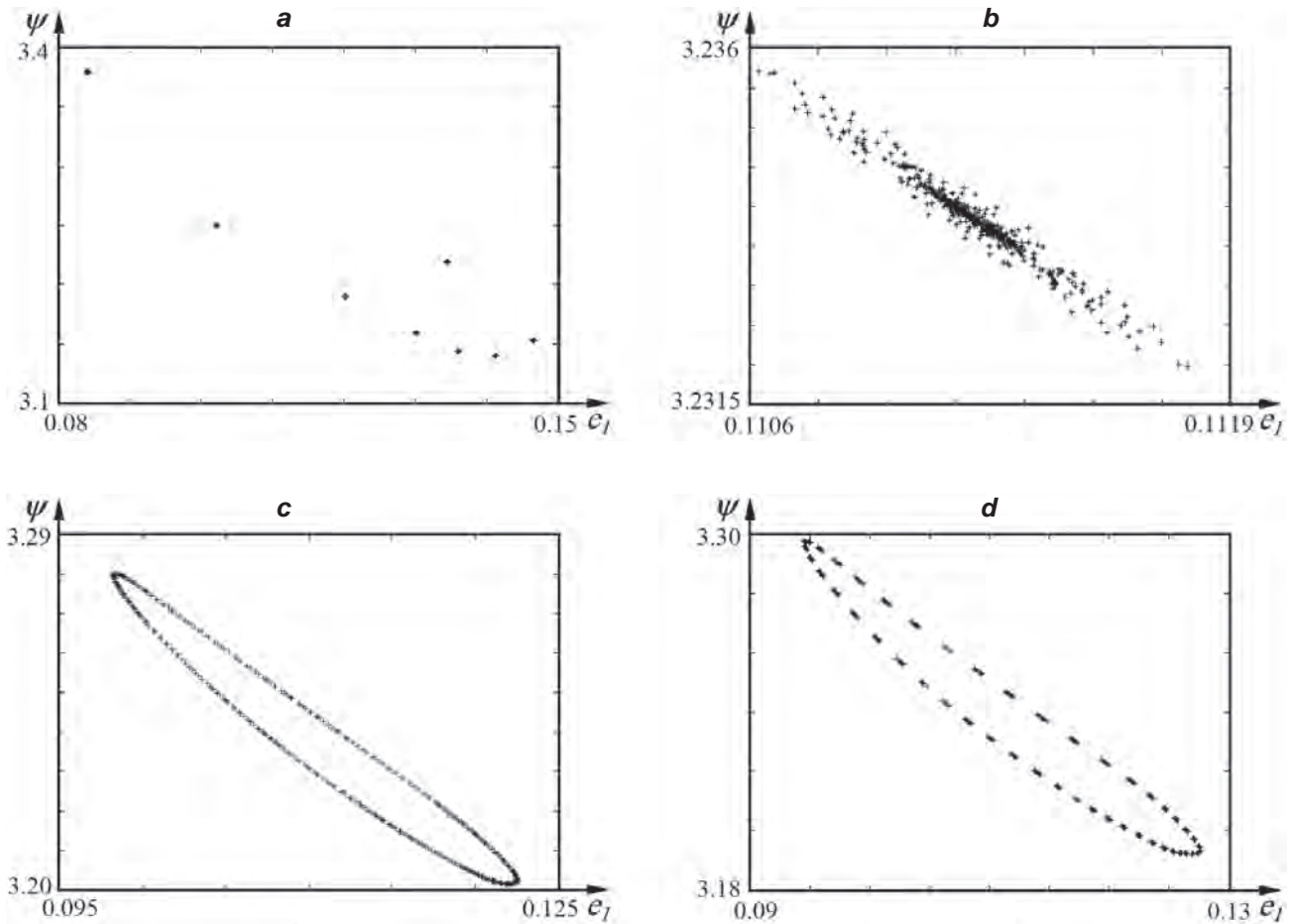

*Fig. 5. Projections of the Poincaré sections on the plane P = 0.2: (a) the regular attractor 8·2⁰ (α = 0.0321107); (b) the regular attractor of a quasiperiodic cycle ≈ 1·2⁰ on a toroidal surface α = 0.0321125; (c) the regular attractor of a quasiperiodic cycle ≈ n·2⁰ on a toroidal surface α = 0.0321286; (d) the regular attractor of a periodic cycle 47·2⁰ on a toroidal surface (α = 0.032135)*

According to Fig. 5,*b,c*, we constructed the Poincaré maps for the variable $e_1$ in Fig. 6,*a,b*. For a quasiperiodic cycle ≈ 1·2⁰, the Poincaré map is the set of points with their maximum localization in the central region. For the regular attractor of a quasiperiodic cycle ≈ n·2⁰, the Poincaré map is an oval characterizing the repetition of oscillations on the torus.

As an example for the visual representation of a complicated structure of the regular attractor 47·2⁰ on a toroidal surface (α = 0.032135), we present its three-dimensional projections in the coordinates ($e_1,\psi,G$) (*a*) and ($B,\psi,O_2$) (*b*) in Fig. 7,*a,b*. Despite a high strokeness, the trajectory returns strictly regularly at any point of the cycle in the time interval strictly equal to the cycle period.

The attractors under consideration are presented in Table. The strange attractor 8·2ˣ (α = 0.03211295) is not unique in the given region. Strange attractors can be found at other values of α at a more thorough study. But all they have a common mechanism of formation of their structure. In accordance with the values of Lyapunov exponents ($\lambda_3-\lambda_{10}$) < 0, the trajectories of the given attractor exponentially converge in the process of evolution. In the same directions, the element of the phase volume shrinks. But since $\lambda_1 > 0$, the trajectories diverge in this specific direction. Moreover, the element of the phase volume is extended in this direction, though it shrinks during the evolution on the whole, $\Lambda < 0$. Such a motion results in that the phase trajectories form folds and mix with one another. In such situations, the smallest fluctuation will cause the randomness of a trajectory.

The elongation and the folding of a chaotic attractor eliminate the initial information and replace it by a new one. Thus, the course of evolution of the system becomes unpredictable. The system can be characterized by the deterministic chaos as a model of chaos arising in the real system. The metabolic process in a cell adapts itself to these conditions, by preserving its stability in certain limits of the stability of a strange attractor.

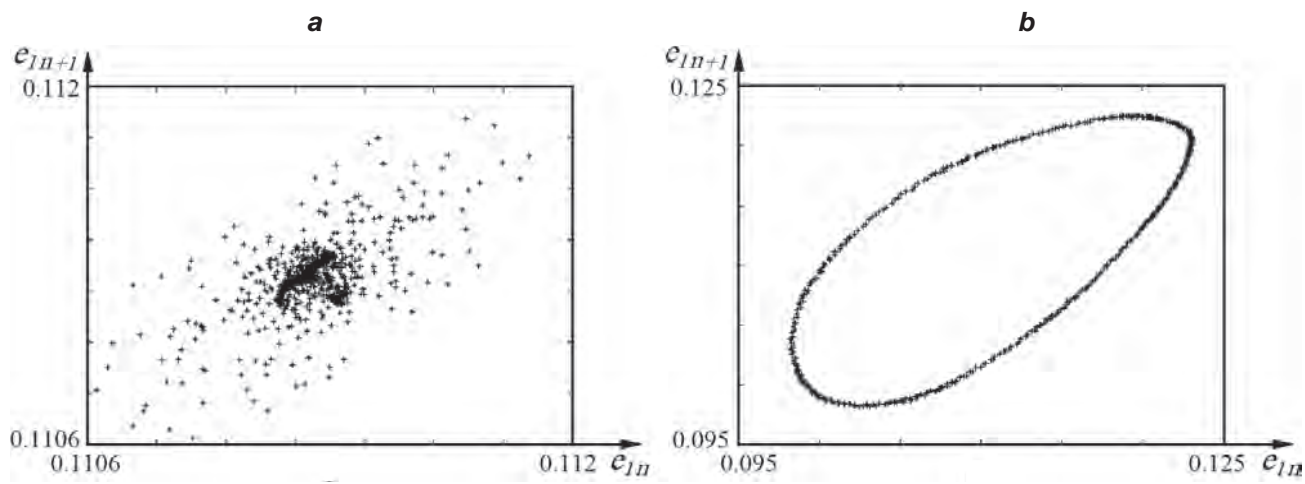

*Fig. 6. Poincaré map for the variable $e_1$: (a) the regular attractor of a quasiperiodic cycle ≈ 1·2⁰ on a toroidal surface (α = 0.0321125); (b) the regular attractor of a quasiperiodic cycle ≈ 47·2⁰ on a toroidal surface (α = 0.0321286)*

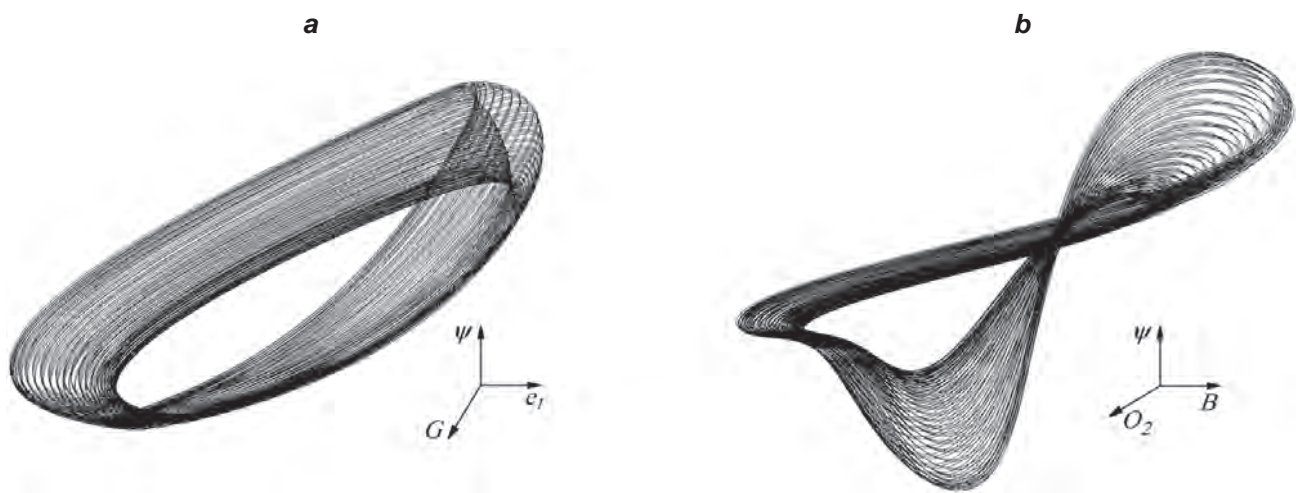

*Fig. 7. (a) Projections of the phase portrait of the regular attractor of a periodic cycle 47·2⁰ on a toroidal surface at α = 0.032135 in the three-dimensional systems of coordinates: $(e_1,\psi,G)$; (b) Projections of the phase portrait of the regular attractor of a periodic cycle 47·2⁰ on a toroidal surface at α = 0.032135 in the three-dimensional systems of coordinates $(B,\psi,O_2)$*

The further complete investigation of modes of the metabolic process running in cells within the given model and the consideration of conditions for the appearance of a deterministic chaos will be performed in the subsequent works.

With the use of the mathematical model of a cell *Arthrobacter globiformis*, we have studied the internal dynamics of its metabolic process in the oscillatory mode. We have found the scenarios of formation and destruction of regular and strange attractors with various periods and types depending on the dissipation of the kinetic membrane potential. The limits of phase-parametric characteristics of the regions, where the bifurcations and the transitions «order-chaos», «chaos-order» and «order-order» arise, are determined. By calculating the total spectra of Lyapunov indices, divergences, Poincaré sections, and Poincaré maps, we studied the regularity of the sequence of attractors appearing on a toroidal surface. In particular, we found a strange attractor, whose structure is formed with the help of a fold. In such a mode, the metabolic process in a cell adapts itself and conserves its stability characteristic of a strange attractor.

Within this model, the application of classical tools of the nonlinear dynamics to the study of the self-organization in metabolic processes of cells with the use of multidimensional nonlinear systems of differential equations has been demonstrated. This allows one to reveal the structural-functional connections in a cell, which underlie the formation of its structure.

The study of various types of the chaotic dynamics inherent in the metabolic processes gives possibility to clarify the basic laws of self-organization in open nonlinear systems of any nature.

*The work is supported by the project N 0113U001093 of the National Academy of Sciences of Ukraine.*

## АВТОКОЛИВАЛЬНА ДИНАМІКА МЕТАБОЛІЧНОГО ПРОЦЕСУ КЛІТИНИ

*В. Й. Грицай[1], І. В. Мусатенко[2]*

[1]Інститут теоретичної фізики
ім. М. М. Боголюбова НАН України, Київ;
e-mail: vgrytsay@bitp.kiev.ua;
[2]Київський національний університет
імені Тараса Шевченка, Україна;
e-mail: ivmusatenko@gmail.com

В роботі досліджується математична модель автоколивальної динаміки в метаболічному процесі клітини. Розраховані фазопараметричні характеристики зміни виду атракторів від величини дисипації кінетичного мембранного потенціалу. Знайдено біфуркації і сценарії переходів: «порядок-хаос», «хаос-порядок» та «порядок-порядок». Побудовані проекції багатомірних фазових портретів атракторів, перерізів та відображень Пуанкаре. Досліджено процес самоорганізації регулярних атракторів через виникнення тора. Розраховані повні спектри експонентів Ляпунова, дивергенції, що характеризують структурну стійкість знайдених атракторів.

Отримані результати демонструють можливість застосування класичних інструментів нелінійної динаміки для дослідження самоорганізації і виникнення хаосу в метаболічних процесах клітин.

К л ю ч о в і  с л о в а: математична модель, метаболічні процеси, самоорганізація, дивний атрактор, переріз Пуанкаре, експоненти Ляпунова.

## АВТОКОЛЕБАТЕЛЬНАЯ ДИНАМИКА МЕТАБОЛИЧЕСКОГО ПРОЦЕССА КЛЕТКИ

*В. И. Грицай[1], И. В. Мусатенко[2]*

[1]Институт теоретической физики
им. Н. Н. Боголюбова НАН Украины, Киев;
e-mail: vgrytsay@bitp.kiev.ua;
[2]Киевский национальный университет
имени Тараса Шевченко, Украина;
e-mail: ivmusatenko@gmail.com

В работе исследуется математическая модель автоколебательной динамики в метаболическом процессе клетки. Рассчитаны фазопараметрические характеристики изменений вида аттракторов от величины диссипации кинетического мембранного потенциала. Найдены бифуркации и сценарии переходов: «порядок-хаос», «хаос-порядок» и «порядок-порядок». Построены проекции многомерных фазовых портретов аттракторов, сечений и отображений Пуанкаре. Исследован процесс самоорганизации регулярных аттракторов вследствие возникновения тора. Рассчитаны полные спектры экспонентов Ляпунова, дивергенции, характеризующит структурную устойчивость найденных аттракторов.

Полученные результаты демонстрируют возможность применения классических инструментов нелинейной динамики для исследования самоорганизации и возникновения хаоса в метаболических процессах клеток.

К л ю ч е в ы е  с л о в а: математическая модель, метаболические процессы, самоорганизация, странный аттрактор, сечение Пуанкаре, экспоненты Ляпунова.